\documentclass[pre,reprint,groupedaddress,nofootinbib]{revtex4-1}

\usepackage{amsmath,amssymb,graphicx,enumerate,siunitx,float,mathtools}
\usepackage{hyperref}
\usepackage[table]{xcolor}
\hypersetup{colorlinks = true, linktocpage = true, bookmarksopen = true, linkcolor = blue, urlcolor=blue, citecolor = blue, allcolors=blue}
\usepackage{array}
\usepackage[margin=1.7cm]{geometry}

\definecolor{tableshade}{RGB}{198.6000  227.1000  239.1000}

\newcommand{\revision}[1]{{\color{black}{#1}}}
\newcommand{\change}[1]{{\color{black}{#1}}}

\graphicspath{{./Figures/}}

\begin{document}

\title{\revision{Finite transition times for multispecies diffusion in heterogeneous media\\ coupled via first-order reaction networks}}

\author{Elliot\hspace{0.1cm}J.\hspace{0.1cm}Carr}
\email[]{elliot.carr@qut.edu.au}
\affiliation{School of Mathematical Sciences, Queensland University of Technology (QUT), Brisbane, Australia} 
\author{Jonah J.\hspace{0.1cm}Klowss}
\affiliation{School of Mathematical Sciences, Queensland University of Technology (QUT), Brisbane, Australia} 

\begin{abstract}
Calculating how long a coupled multi-species reactive-diffusive transport process in a heterogeneous medium takes to effectively reach steady state is important in many applications. In this paper, we show how the time required for such processes to transition to within a small specified tolerance of steady state can be calculated accurately without having to solve the governing time-dependent model equations. \revision{Our approach is valid for general first-order reaction networks and an arbitrary number of species. Three} numerical examples are presented to confirm the analysis and investigate the efficacy of the approach. A key finding is that for sequential reactions our approach works better provided the two smallest reaction rates are well separated. MATLAB code implementing the methodology and reproducing the results in the paper is made available.
\end{abstract}

\pacs{}

\maketitle

\section{Introduction}
Calculating how long a transport process takes to complete is important in many diverse modelling applications including drug delivery \cite{carr_2019a,simon_2009,simon_2016}, biological tissue development \cite{ellery_2012b}, transport in groundwater aquifers \cite{carr_2018a,ellery_2012b,simpson_2013a} and various heat conduction applications \cite{hickson_2009_part1,rusagara_2014}. Unfortunately, partial differential equation descriptions of such processes give rise to the impractical answer of an infinite completion time, that is, an infinite amount of time is required for the transient solution of the governing equations to transition from initial to steady state. This has led to the introduction of the concept of a \textit{finite transition time} (depending on the application other names include response time \cite{carr_2018a}, critical time \cite{ellery_2012b} and release time \cite{carr_2019a}), which provides the time required for the process to ``effectively'' reach steady state, that is, within a small prescribed tolerance. 

The concept of a finite measure of the time to reach steady state was pioneered by McNabb and colleagues \cite{mcnabb_1991a,mcnabb_1991b} with the introduction of the mean action time. Here, the transition from initial to steady-state is defined in terms of a cumulative distribution function with the mean or first moment used as an estimate of the finite transition time. The attraction of this approach is that the first moment (and indeed higher-order moments) can be computed without solving the underlying governing model equations. Over the past two decades, many authors have built on McNabb and Wake's original idea with a primary focus on diffusion problems and low accuracy ``rule-of-thumb'' estimates \cite{mcnabb_1991b} based on lower-order moments \cite{landman_2000,berezhkovskii_2010,jazaei_2014,ellery_2012a,ellery_2012b,carr_2017b,carr_2018b,simpson_2013a}. An extensive review of this work can be found elsewhere \cite{carr_2017b}.  

More recently, \citet{carr_2017b} demonstrated how finite transition times for diffusion processes can be accurately calculated using high-order moments resulting in high accuracy estimates without have to solve the underlying governing model equations. This work has since been extended to a heterogeneous diffusion problem arising in groundwater modelling applications \cite{carr_2018a} and multi-layer diffusion problems in spherical coordinates arising in drug release applications \cite{carr_2019a}. In this paper, we extend this analysis to coupled multi-species reaction-diffusion models demonstrating how the finite transition time for each individual species can be accurately calculated. Our work finds practical application to numerous problems involving coupled reaction-diffusion models such as reactive contaminant transport in groundwater systems \cite{sun_1999b} and the diffusion and proliferation of different generations within a population of cells \cite{simpson_2015}.

The remaining sections of the paper are organised as follows. In the next section, we describe the coupled $n$-species reaction-diffusion model considered in this work. In section \ref{sec:finite_transition_times}, we study the spatially-discretised form of the reaction-diffusion model to derive an asymptotic estimate of the finite transition time for each species in terms of the aforementioned moments. In section \ref{sec:moments} we explain how the moments (and hence the finite transition time) can be calculated for each of the $n$ species without having to solve the governing model equations. Results are then presented in section \ref{sec:results} verifying the analysis and investigating the accuracy of the proposed approach. Finally, the limitations and conclusions of the work are discussed and summarised in section \ref{sec:conclusions}.

\section{Coupled reaction-diffusion model}
\label{sec:model}
We consider $n$-species reactive-diffusive transport in a heterogeneous medium \revision{coupled via general first order reactions} \cite{sun_1999b,simpson_2015,simpson_2013a}:
\begin{align}
\label{eq:pde1}
\frac{\partial c_{1}}{\partial t} &= \frac{\partial}{\partial x}\left(D(x)\frac{\partial c_{1}}{\partial x}\right) + \revision{\sum_{\ell=1}^{n} \mu_{1,\ell}c_{\ell}},\\
\frac{\partial c_{2}}{\partial t} &= \frac{\partial}{\partial x}\left(D(x)\frac{\partial c_{2}}{\partial x}\right) + \revision{\sum_{\ell=1}^{n} \mu_{2,\ell}c_{\ell}},\\
&\hspace{0.2cm}\vdots\nonumber\\
\label{eq:pden}
\frac{\partial c_{n}}{\partial t} &= \frac{\partial}{\partial x}\left(D(x)\frac{\partial c_{n}}{\partial x}\right) + \revision{\sum_{\ell=1}^{n} \mu_{n,\ell}c_{\ell}},
\end{align}
where $c_{i}(x,t)$ is the concentration of species $i$ at position $x\in(0,L)$ and time $t>0$, $D(x)$ is the spatially-dependent (heterogeneous) diffusivity and \revision{$\mu_{i,\ell}$ is a first-order reaction rate representing the decay of species $i$ ($i = \ell$, $\mu_{i,i} < 0$) or the production of species $i$ from species $\ell$ ($i\neq \ell$, $\mu_{i,\ell} > 0$)}. The coupled reaction-diffusion equations are paired with the following initial and boundary conditions: 
\begin{gather}
c_{i}(x,0) = 0,\enspace\text{for all $i = 1,\hdots,n$,}\\
\label{eq:bcL}
c_{i}(0,t) = \revision{c_{b,i},\enspace\text{for all $i = 1,\hdots,n$,}}\\
\label{eq:bcR}
\frac{\partial c_{i}}{\partial x}(L,t) = 0,\enspace\text{for all $i = 1,\hdots,n$},
\end{gather}
where \revision{$c_{b,i}$ is the specified concentration of species $i$ at $x = 0$}. While we consider only the above boundary conditions in our analysis, our approach is easily modified for other types of boundary conditions (e.g. flux-specified at $x = 0$) as discussed elsewhere \cite{carr_2017b}.

The steady-state solution of the coupled reaction-diffusion model (\ref{eq:pde1})--(\ref{eq:bcR}), $c_{i,\infty}(x) := \lim_{t\rightarrow\infty} c_{i}(x,t)$ for all $i = 1,\hdots,n$, satisfies:
\begin{align}
\label{eq:pde_ss}
0 &= \frac{\text{d}}{\text{d}x}\left(D(x)\frac{\text{d} c_{1,\infty}}{\text{d} x}\right) + \revision{\sum_{\ell=1}^{n} \mu_{1,\ell}c_{\ell,\infty}},\\
0 &= \frac{\text{d}}{\text{d}x}\left(D(x)\frac{\text{d} c_{2,\infty}}{\text{d} x}\right) + \revision{\sum_{\ell=1}^{n} \mu_{2,\ell}c_{\ell,\infty}},\\
&\hspace{0.2cm}\vdots\nonumber\\
0 &= \frac{\text{d}}{\text{d}x}\left(D(x)\frac{\text{d} c_{n,\infty}}{\text{d} x}\right) + \revision{\sum_{\ell=1}^{n} \mu_{n,\ell}c_{\ell,\infty}},
\end{align}
subject to the boundary conditions:
\begin{gather}
\label{eq:bcL_ss}
c_{i,\infty}(0) = \revision{c_{b,i},\enspace\text{for all $i = 1,\hdots,n$,}}\\
\label{eq:bcR_ss}
\frac{\text{d}c_{i,\infty}}{\text{d}x}(L) = 0,\enspace\text{for all $i = 1,\hdots,n$}.
\end{gather}

\section{Finite transition times}
\label{sec:finite_transition_times}
We define the finite transition time for species $i$ at position $x$ as the time $\tau_{i} >0$ satisfying the equation:
\begin{align}
\label{eq:taui_definition}
f_{i}(\tau_{i}; x) = 1-\delta,
\end{align}
where $0<\delta\ll 1$ is a small specified tolerance and
\begin{align}
\label{eq:fi}
f_{i}(t; x) = 1 - \left[\frac{c_{i}(x,t) - c_{i,\infty}(x)}{c_{i,0}(x) - c_{i,\infty}(x)}\right],
\end{align}
\revision{with $c_{i,0}(x)$ used to denote $c_{i}(x,0)$.} Note that $f_{i}(0; x) = 0$ and $\lim_{t\rightarrow\infty}f_{i}(t; x) = 1$. Provided $c_{i}(x,t)$ transitions from initial to steady state monotonically, $f_{i}(t; x)$ is a cumulative distribution function \cite{ellery_2012b} and specifies the fraction of the transition from initial to steady state completed at time $t$. We also remark that it is clear from Eqs (\ref{eq:taui_definition})--(\ref{eq:fi}) that $\tau_{i}$ is a function of $x$. This makes sense since different spatial locations take different amounts of time to transition from initial to steady state. As we have done in Eq (\ref{eq:taui_definition}), when it is convenient to do so we will drop this dependence on $x$ and write $\tau_{i}$.

As previously mentioned, the goal of this paper is to show how the transition time $\tau_{i}$ for each species $i$ can be estimated accurately without requiring the transient solutions $c_{i}(x,t)$ ($i=1,\hdots,n$) of the coupled reaction-diffusion model (\ref{eq:pde1})--(\ref{eq:bcR}). We achieve this by first studying the spatially-discrete form of the governing model equations as follows.

Consider a uniform grid on the interval $[0,L]$ consisting of $N$ nodes: $0 = x_{1} < x_{2} < \cdots < x_{N} = L$ where $x_{j} = (j-1)h$ with grid spacing $h = L/(N-1)$. Let $c_{i,j}(t)$ be the numerical approximation to $c_{i}(x_{j},t)$. Applying a standard discretisation method (e.g. finite volume method) to discretise the governing equations (\ref{eq:pde1})--(\ref{eq:bcR}) in space, yields an initial value problem involving a system of linear differential equations of dimension $\mathcal{N} = n(N-1)$, expressible in matrix form as follows: 
\begin{gather}
\label{eq:ode_system}
\frac{\text{d}\mathbf{c}}{\text{d}t} = -\mathbf{A}\mathbf{c} + \mathbf{b},\quad\mathbf{c}(0)=\mathbf{0},
\end{gather}
where $\mathbf{A}\in\mathbb{R}^{\mathcal{N}\times \mathcal{N}}$, $\mathbf{0}\in\mathbb{R}^{\mathcal{N}\times 1}$ is a zero vector, $\mathbf{b}\in\mathbb{R}^{\mathcal{N}\times 1}$ and 
\begin{gather*}
\mathbf{c} = \left[\begin{matrix} \mathbf{c}_{1}\\ \mathbf{c}_{2}\\ \vdots\\ \mathbf{c}_{n}\end{matrix}\right]\in\mathbb{R}^{\mathcal{N}\times 1}.
\end{gather*}
In the above equation, $\mathbf{c}_{i} = (c_{i,2},\hdots,c_{i,N})^{T}\in\mathbb{R}^{(N-1)\times 1}$ for all $i = 1,\hdots,n$. Note that discrete values $c_{i,1}$ ($i=1,\hdots,n$) are excluded from the system (\ref{eq:ode_system}) since they are known from the boundary condition (\ref{eq:bcL}): $c_{i,1} = c_{i}(0,t)$. 

The exact solution of the differential system (\ref{eq:ode_system}) is
\begin{gather*}
\mathbf{c}(t) = \mathbf{c}_{\infty} + e^{-t\mathbf{A}}\left(\mathbf{c}_{0} - \mathbf{c}_{\infty}\right),
\end{gather*}
where $\mathbf{c}_{\infty} = \mathbf{A}^{-1}\mathbf{b}$ is the steady-state solution. Assume the eigenvalues of $\mathbf{A}$ are real, positive, distinct and ordered such that $\lambda_{1} < \lambda_{2} < \cdots < \lambda_{\mathcal{N}}$ and note from the definition of the finite transition time (\ref{eq:taui_definition})--(\ref{eq:fi}) that smaller values of $\delta$ translate to larger values of $\tau_{i}$.  Provided $e^{-t\lambda_{1}}\gg e^{-t\lambda_{2}}$ (i.e. $e^{-t(\lambda_{1}-\lambda_{2})}\gg 1$), we deduce from previous work \cite{carr_2018a} that $c_{i}(x,t)$ has the following functional form:
\begin{align}
\label{eq:ci_larget}
c_{i}(x,t) \approx c_{i,\infty}(x) + v_{i}(x)e^{-t\lambda_{1}},
\end{align}
for large $t$\revision{, where $v_{i}$ is some function of $x$.} It follows from (\ref{eq:fi}) that $f_{i}(t; x)$ has the following functional form:
\begin{align}
\label{eq:fi_larget}
f_{i}(t; x) \approx 1 - \alpha_{i}(x)e^{-t\lambda_{1}},
\end{align}
for large $t$, where \revision{$\alpha_{i}(x) = v_{i}(x)/[c_{i,0}(x) - c_{i,\infty}(x)]$}. Combining (\ref{eq:fi_larget}) with (\ref{eq:taui_definition}) and solving for $\tau_{i}$ yields the following estimate of the finite transition time for species $i$:
\begin{gather}
\label{eq:transition_time}
\tau_{i}(x) \approx \frac{1}{\lambda_{1}}\log\left(\frac{\alpha_{i}(x)}{\delta}\right).
\end{gather} 
Define the $k$th temporal moment for species $i$ as follows
\begin{gather}
\label{eq:Mik}
M_{i,k}(x) = \int_{0}^{\infty} t^{k}f_{i}'(t; x)\,\text{d}t,
\end{gather}
where $k = 1,\hdots,m$, 
\begin{align}
\label{eq:fi_dash}
f_{i}'(t; x) = -\frac{\partial}{\partial t}\left[\frac{c_{i}(x,t) - c_{i,\infty}(x)}{c_{i,0}(x)-c_{i,\infty}(x)}\right],
\end{align}
and $m$ is a specified positive integer. Note that $M_{i,k}(x) > 0$ since $f_{i}'(t; x)>0$ for all $t > 0$; the latter deduced from Eq (\ref{eq:fi_dash}) and the monotonicity of $c_{i}(x,t)$. Following previous work \cite{carr_2017b,carr_2018a}, one can show for the reaction-diffusion model (\ref{eq:pde1})--(\ref{eq:bcR}) the unknown parameters in the finite transition time estimate (\ref{eq:transition_time}) can be approximated as follows:
\begin{gather}
\label{eq:lambda1}
\lambda_{1}(x) \approx \frac{kM_{i,k-1}(x)}{M_{i,k}(x)},\quad\\
\label{eq:alpha1}
\alpha_{1}(x) \approx \frac{M_{i,k}(x)}{k!}\left(\frac{kM_{i,k-1}(x)}{M_{i,k}(x)}\right)^{k},
\end{gather}
for large $k$. In summary, we have the following formula for the finite transition time for species $i$:
\begin{gather}
\label{eq:tau_x}
\tau_{i}(x) \approx \frac{M_{i,k}(x)}{kM_{i,k-1}(x)}\log\left[\frac{M_{i,k}(x)}{k!\,\delta}\left(\frac{kM_{i,k-1}(x)}{M_{i,k}(x)}\right)^{k}\right],
\end{gather}
with the expectation of increasing accuracy for increasing values of $k$ \cite{carr_2017b} due to use of the approximations (\ref{eq:lambda1})--(\ref{eq:alpha1}).

\section{Temporal Moments}
\label{sec:moments}
The key attraction of the finite transition time formula (\ref{eq:tau_x}) is that the moments can be calculated without calculating the transient solution $c_{i}(x,t)$ \cite{carr_2017b}. This is achieved for the coupled $n$-species reaction-diffusion model (\ref{eq:pde1})--(\ref{eq:bcR}) by extending analysis previously presented for single-species diffusion \cite{carr_2017b,carr_2018a}. We remark that an alternative method for deriving the first two moments is provided by \citet{simpson_2013a} for the special case of $n = 2$ species.

First note that for each species $i$, the reaction-diffusion equations (\ref{eq:pde1})--(\ref{eq:pden}) are expressible in the general form:
\begin{gather}
\label{eq:pdei}
\frac{\partial c_{i}}{\partial t} = \frac{\partial}{\partial x}\left(D(x)\frac{\partial c_{i}}{\partial x}\right) + \revision{\sum_{\ell=1}^{n}\mu_{i,\ell}c_{\ell}},
\end{gather}
Applying integration by parts to the integral (\ref{eq:Mik}) yields:
\begin{gather}
\label{eq:Mik_2}
M_{i,k}(x) = k\int_{0}^{\infty} t^{k-1}\left[\frac{c_{i}(x,t)-c_{i,\infty}(x)}{c_{i,0}(x) - c_{i,\infty}(x)}\right]\,\text{d}t,
\end{gather}
since $c_{i}(x,t)-c_{i,\infty}(x)$ approaches zero faster than $t^{k}$ approaches $\infty$ \cite{ellery_2012b,simpson_2013a,carr_2017b}. Following standard convention \cite{carr_2017b,carr_2018b}, we define 
\begin{gather}
\label{eq:Mikbar}
\overline{M}_{i,k}(x) = M_{i,k}(x)\left[c_{i,\infty}(x)-c_{i,0}(x)\right],
\end{gather}
or equivalently:
\begin{align}
\label{eq:Mikbar1}
\overline{M}_{i,k}(x) &= \int_{0}^{\infty} t^{k}\frac{\partial}{\partial t}\left[c_{i}(x,t)-c_{i,\infty}(x)\right]\,\text{d}t,\\
\label{eq:Mikbar2}
\overline{M}_{i,k}(x) &= k\int_{0}^{\infty} t^{k-1}\left[c_{i,\infty}(x)-c_{i}(x,t)\right]\,\text{d}t,
\end{align}
when using Eqs (\ref{eq:Mik})--(\ref{eq:fi_dash}) and (\ref{eq:Mik_2}), respectively.

Applying the linear operator $\mathcal{L} = \frac{\partial}{\partial x}\left(D(x)\frac{\partial}{\partial x}\right)$ to both sides of (\ref{eq:Mikbar2}) and making use of the reaction-diffusion equation (\ref{eq:pdei}) and its steady-state analogue yields:
\begin{align*}
\mathcal{L}\overline{M}_{i,k}(x) &= k\int_{0}^{\infty} t^{k-1}\frac{\partial}{\partial t}\left[c_{i,\infty}(x)-c_{i}(x,t)\right]\,\text{d}t\\ 
&\quad\revision{- \sum_{\ell=1}^{n}\mu_{i,\ell}\left(k\int_{0}^{\infty} t^{k-1} \left[c_{\ell,\infty}(x)-c_{\ell}(x,t)\right]\,\text{d}t\right)}.
\end{align*}
Combining the above equation with the expressions for $\overline{M}_{i,k}$ (\ref{eq:Mikbar1})--(\ref{eq:Mikbar2}) yields the differential equation:
\begin{align*}
\frac{\text{d}}{\text{d}x}\left(D(x)\frac{\text{d}\overline{M}_{i,k}}{\text{d}x}\right) = - k\overline{M}_{i,k-1} \revision{-\sum_{\ell=1}^{n}\mu_{i,\ell}\overline{M}_{\ell,k}}.
\end{align*}
Hence, the functions $\overline{M}_{i,k}$ ($i = 1,\hdots,n$) satisfy a coupled system of differential equations:
\begin{align}
\label{eq:moment_de1}
\frac{\text{d}}{\text{d}x}\left(D(x)\frac{\text{d}\overline{M}_{1,k}}{\text{d}x}\right)  \revision{+\sum_{\ell=1}^{n}\mu_{1,\ell}\overline{M}_{\ell,k}} &= - k\overline{M}_{1,k-1},\\
\frac{\text{d}}{\text{d}x}\left(D(x)\frac{\text{d}\overline{M}_{2,k}}{\text{d}x}\right) \revision{+\sum_{\ell=1}^{n}\mu_{2,\ell}\overline{M}_{\ell,k}} &= - k\overline{M}_{2,k-1},\\
&\hspace{0.2cm}\vdots\nonumber\\
\label{eq:moment_den}
\frac{\text{d}}{\text{d}x}\left(D(x)\frac{\text{d}\overline{M}_{n,k}}{\text{d}x}\right) \revision{+\sum_{\ell=1}^{n}\mu_{n,\ell}\overline{M}_{\ell,k}} &= - k\overline{M}_{n,k-1}.
\end{align}
The appropriate boundary conditions are:
\begin{gather}
\label{eq:moment_bcL}
\overline{M}_{i,k}(0) = 0,\enspace\text{for all $i = 1,\hdots,n$},\\
\label{eq:moment_bcR}
\frac{\text{d}\overline{M}_{i,k}}{\text{d}x}(L) = 0,\enspace\text{for all $i = 1,\hdots,n$},
\end{gather}
which are derived by combining the boundary conditions of the reaction-diffusion model (\ref{eq:bcL})--(\ref{eq:bcR}) and its steady-state analogue (\ref{eq:bcL_ss})--(\ref{eq:bcR_ss}) with the form of $\overline{M}_{i,k}(x)$ (\ref{eq:Mikbar2}). 

We solve the boundary value problem (\ref{eq:moment_de1})--(\ref{eq:moment_bcR}) for $\overline{M}_{i,k}$ numerically. Consider a uniform grid on the interval $[0,L]$ consisting of $N$ nodes: $0 = x_{1} < x_{2} < \cdots < x_{N} = L$ where $x_{j} = (j-1)h$ with grid spacing $h = L/(N-1)$. Let $\overline{M}_{i,k,j}$ be the numerical approximation to $\overline{M}_{i,k}(x_{j})$ with $\overline{M}_{i,k,1} = 0$ from Eq (\ref{eq:moment_bcL}). Applying a finite volume method to discretise the governing equations (\ref{eq:moment_de1})--(\ref{eq:moment_bcR}) in space, yields a system of linear equations of dimension $\change{\overline{\mathcal{N}} = nN}$, expressible in the following matrix form:
\begin{gather}
\label{eq:moment_ls}
\change{\widehat{\mathbf{A}}}\,\overline{\mathbf{M}}_{k} = \change{\mathbf{b}_{k-1}}.
\end{gather}
\change{where $\widehat{\mathbf{A}}$ is closely related to $\mathbf{A}$ in (\ref{eq:ode_system}) and $\mathbf{b}_{k-1}$ depends on $\overline{\mathbf{M}}_{k-1}$. The increase in dimension from $\mathcal{N}$ in (\ref{eq:ode_system}) to $\overline{\mathcal{N}}$ in (\ref{eq:moment_ls}) is due to the inclusion of the left boundary condition (\ref{eq:moment_bcL}) in the discrete system (\ref{eq:moment_ls}).} Full details on our finite volume discretisation can be found in our MATLAB code available on GitHub: \href{https://github.com/elliotcarr/Carr2020b}{https://github.com/elliotcarr/Carr2020b}. In the above system, $\overline{\mathbf{M}}_{k}$ is the following $\overline{\mathcal{N}}$ dimensional column vector:
\begin{gather}
\label{eq:moment_Mkvec}
\overline{\mathbf{M}}_{k} = \left[\begin{matrix} \overline{\mathbf{M}}_{1,k}\\ \overline{\mathbf{M}}_{2,k}\\ \vdots\\ \overline{\mathbf{M}}_{n,k}\end{matrix}\right]\in\mathbb{R}^{\change{\overline{\mathcal{N}}}},
\end{gather}
where 
\begin{gather}
\label{eq:moment_Mikvec}
\overline{\mathbf{M}}_{i,k} = \left[\begin{matrix} \overline{M}_{i,k,\change{1}}\\ \overline{M}_{i,k,\change{2}}\\ \vdots\\\overline{M}_{i,k,N}\end{matrix}\right]\in\mathbb{R}^{\change{N}}.
\end{gather} 
With $\overline{\mathbf{M}}_{0}$ known since $\overline{M}_{i,0,j} = c_{i,\infty}(x_{j}) - c_{i,0}(x_{j})$ for all $j = \change{1},\hdots,N$, the higher-order moments can be computed by solving the linear system (\ref{eq:moment_ls}) sequentially:
\begin{align*}
\overline{\mathbf{M}}_{1} &= \change{\widehat{\mathbf{A}}^{-1}\mathbf{b}_{0}},\\
\overline{\mathbf{M}}_{2} &= \change{\widehat{\mathbf{A}}^{-1}\mathbf{b}_{1}},\\
&\hspace{0.2cm}\vdots\nonumber\\
\overline{\mathbf{M}}_{m} &= \change{\widehat{\mathbf{A}}^{-1}\mathbf{b}_{m-1}}.
\end{align*}
For each $k$, the values of $\overline{M}_{i,k,j}$ for all $i = 1,\hdots,n$ and $j = \change{1},\hdots,N$ are then identified from the entries of $\overline{\mathbf{M}}_{k}$ according to equations (\ref{eq:moment_Mkvec})--(\ref{eq:moment_Mikvec}). Finally, combining the approximation
\begin{align*}
M_{i,k}(x_{j}) \approx M_{i,k,j} = \frac{\overline{M}_{i,k,j}}{c_{i,\infty}(x_{j})-c_{i,0}(x_{j})},
\end{align*}
arising from (\ref{eq:Mikbar}) with (\ref{eq:tau_x}) yields the following finite transition time estimate for species $i$ at node $j$:
\begin{gather}
\label{eq:tauixj}
\tau_{i}(x_{j}) \approx \frac{M_{i,k,j}}{kM_{i,k-1,j}}\log\left[\frac{M_{i,k,j}}{k!\,\delta}\left(\frac{kM_{i,k-1,j}}{M_{i,k,j}}\right)^{k}\right] =: \tau_{i,j}.
\end{gather}

\section{Results and discussion}
\label{sec:results}
\begin{figure*}[p]
\includegraphics[width=0.32\textwidth]{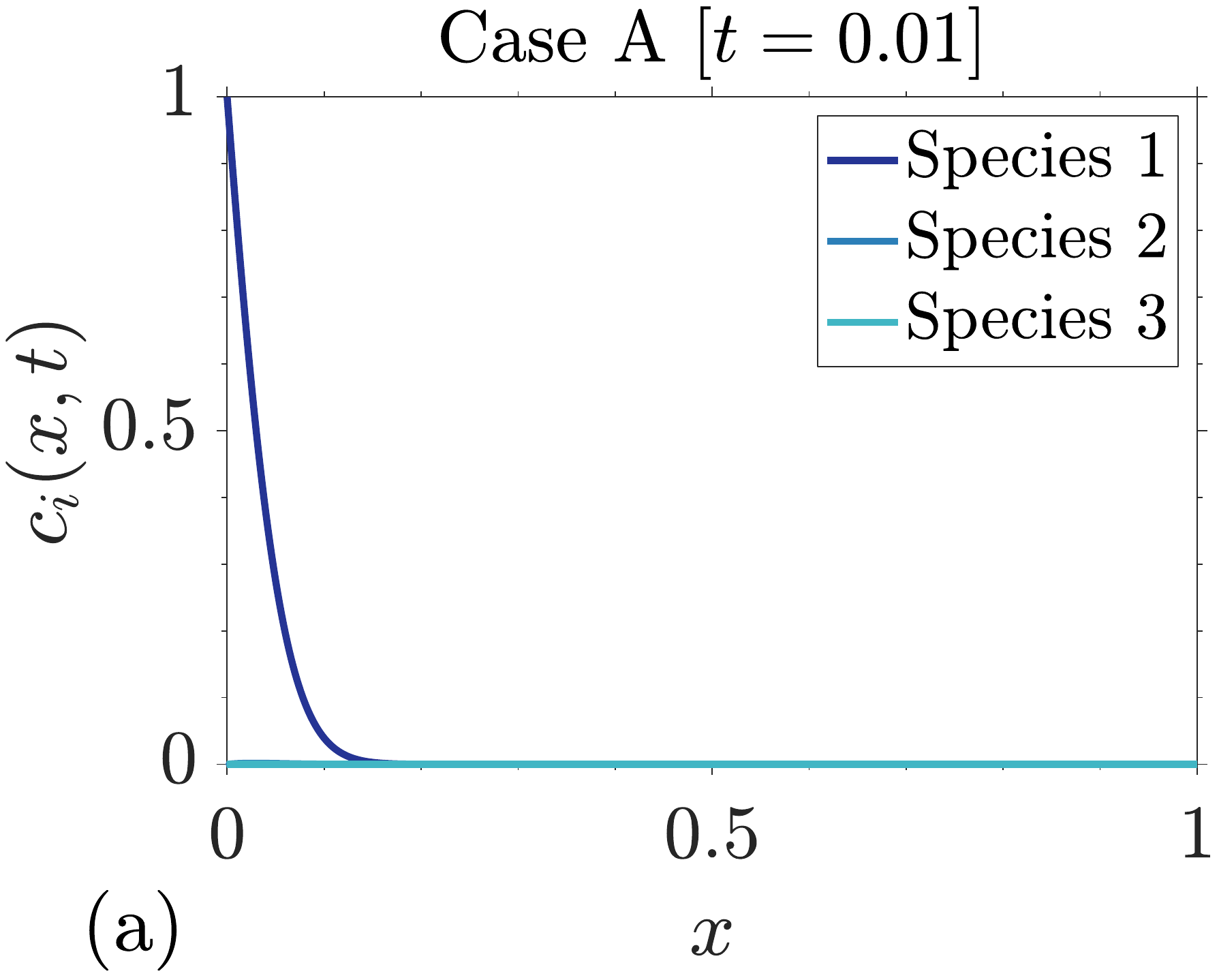}\hspace{0.01\textwidth}
\includegraphics[width=0.32\textwidth]{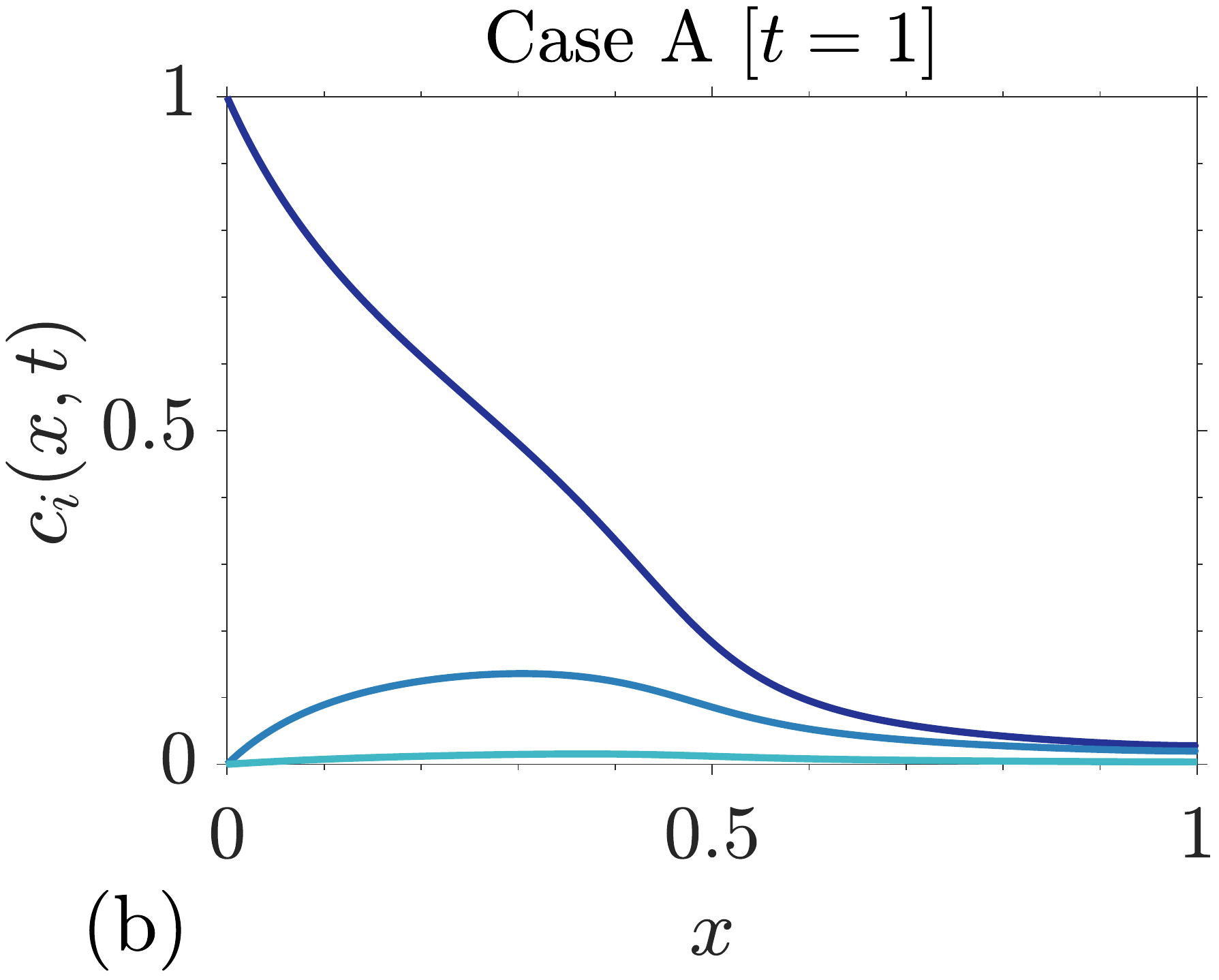}\hspace{0.01\textwidth}\includegraphics[width=0.32\textwidth]{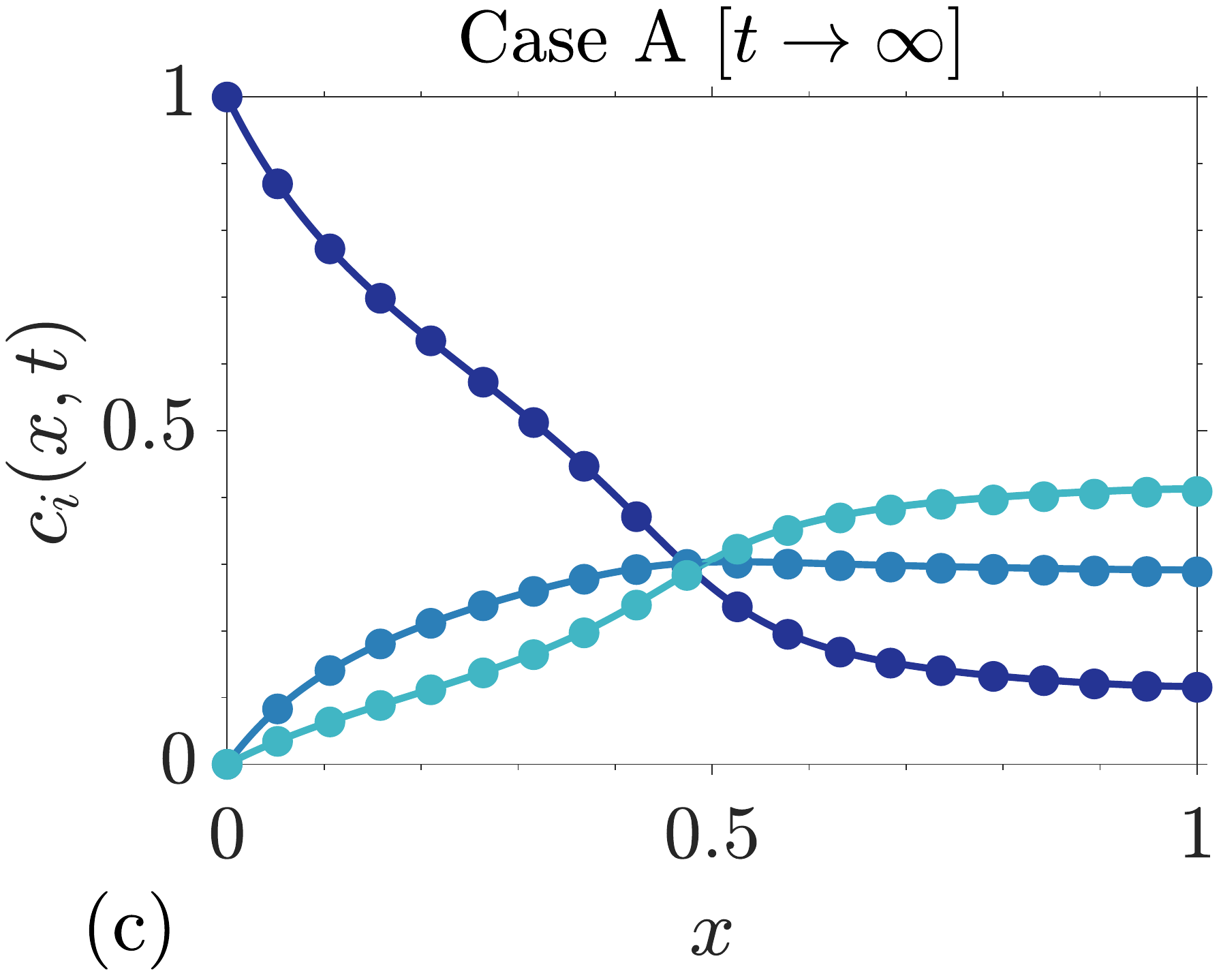}\\\includegraphics[width=0.32\textwidth]{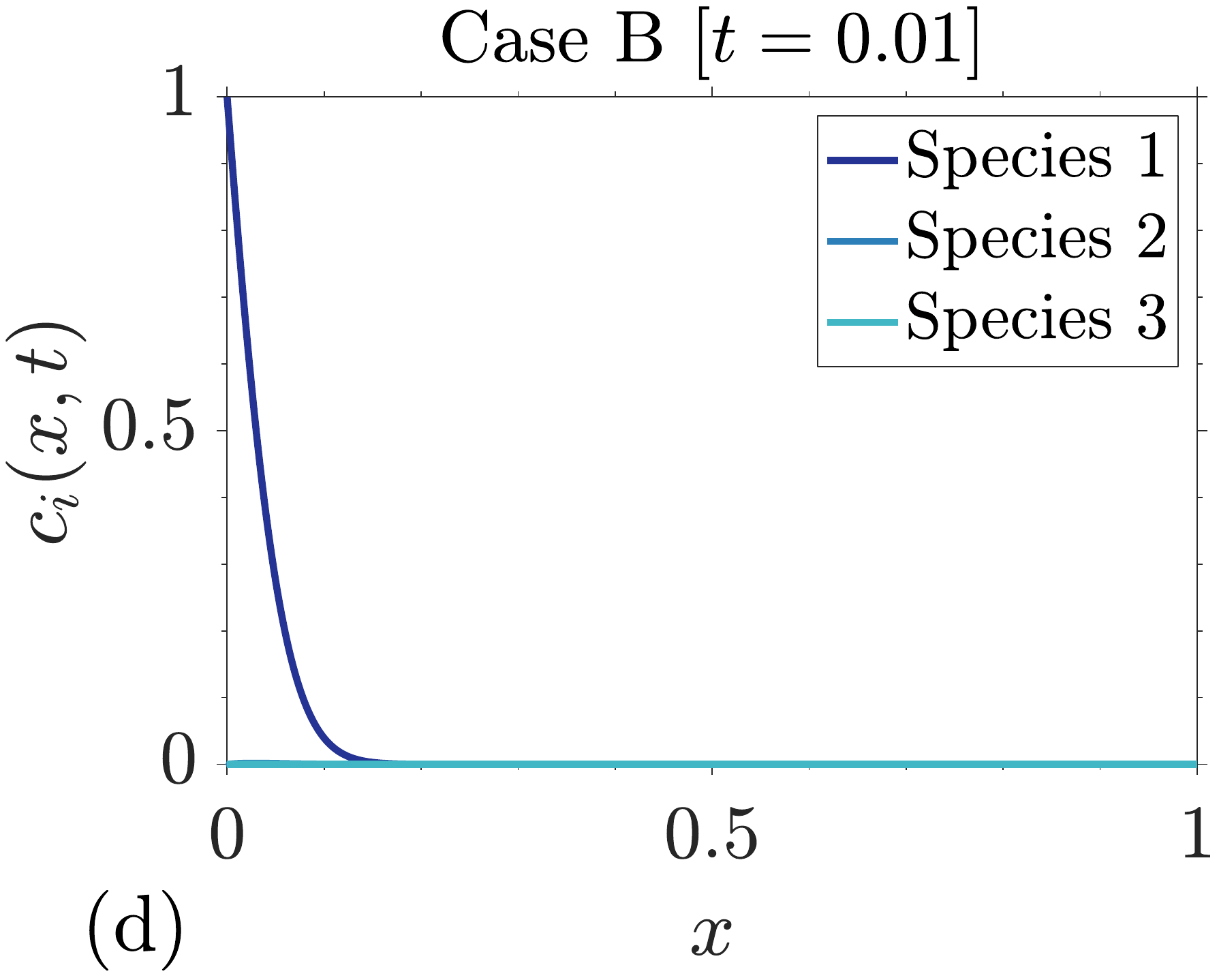}\hspace{0.01\textwidth}
\includegraphics[width=0.32\textwidth]{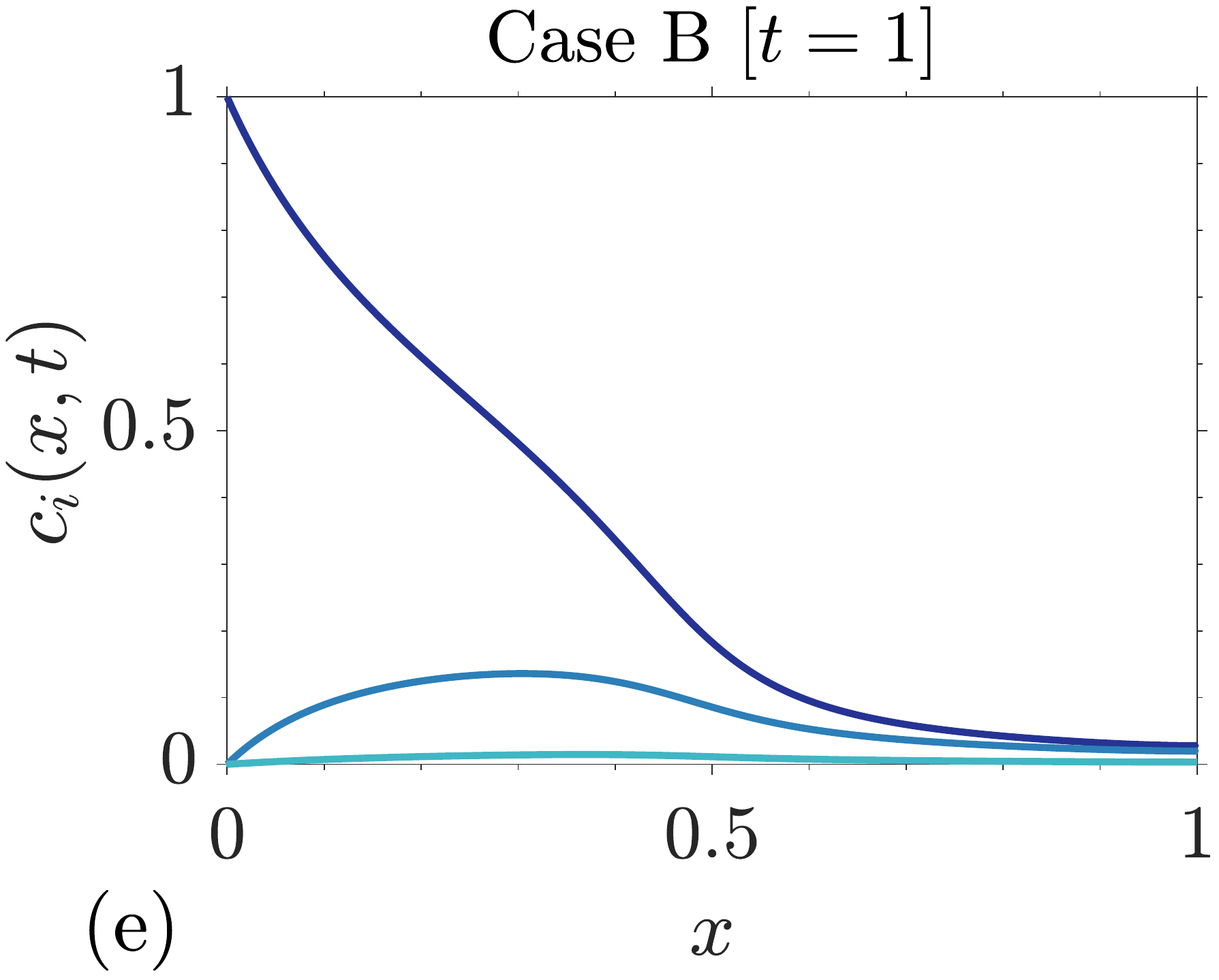}\hspace{0.01\textwidth}\includegraphics[width=0.32\textwidth]{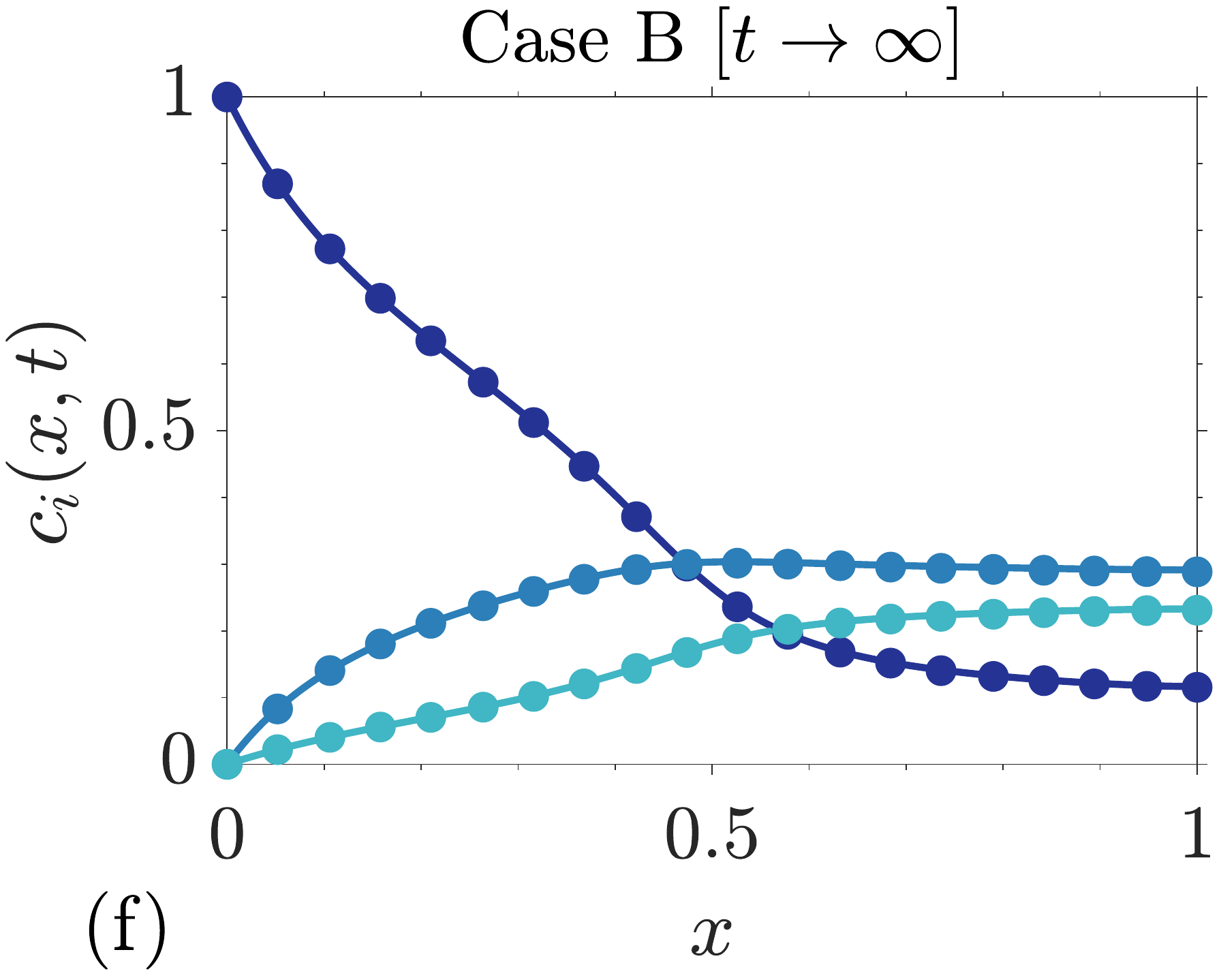}
\caption{Spatial distribution of the species concentration ($c_{i}(x,t)$, $i = 1,2,3$) over time for Cases A and B. Circle markers in the $t\rightarrow\infty$ plots provide the concentration at the following finite transition times: $t = 5.30$ (species 1), $t = 9.05$ (species 2) and $t = 17.16$ (species 3) for Case A and $t = 5.30$ (species 1), $t=9.05$ (species 2) and $t=12.57$ (species 3) for Case B. These finite transition times are calculated at the right boundary ($x = L$) via the formula (\ref{eq:tauixj}) with $j = N$, $\delta = 10^{-2}$ and $k = 15$. All solutions are obtained by solving the coupled reaction-diffusion model (\ref{eq:pde1})--(\ref{eq:bcR}) using the finite volume scheme discussed in Section \ref{sec:results} with $N = 501$ nodes.}
\label{fig:results}
\end{figure*}

\revision{We first apply the finite transition time formula to two test cases involving sequential reactions:
\begin{gather*}
\left[\begin{matrix*}[r] \mu_{1,1} & \mu_{1,2} & \mu_{1,3}\\ \mu_{2,1} & \mu_{2,2} & \mu_{2,3}\\ \mu_{3,1} & \mu_{3,2} & \mu_{3,3}\end{matrix*}\right] := \left[\begin{matrix*}[r] -\mu_{1} & 0 & 0\\ \mu_{1} & -\mu_{2} & 0\\ 0 & \mu_{2} & -\mu_{3}\end{matrix*}\right],
\end{gather*}
where species $1$ produces species $2$ and species $2$ produces species $3$.} Case A considers coupled reactive-diffusive transport in a heterogeneous medium of length $L = 1$ with $n = 3$ species, diffusivity $D(x) = 0.1 + 0.05\sin(10x)$, \revision{$c_{b,i} = 1$ ($i = 1$) or $c_{b,i} = 0$ ($i = 2,3$)}, and reaction rates $\mu_{1} = 0.8$, $\mu_{2} = 0.4$ and $\mu_{3} = 0.1$. Case B is the same as Case A except $\mu_{3} = 0.35$. In Figure \ref{fig:results} we depict the transition from initial to steady state for both test cases. Concentration profiles at the finite transition times shown in the $t\rightarrow\infty$ plots, which are calculated at the right boundary ($x = L$) via the formula (\ref{eq:tauixj}) with $j = N$ and $\delta = 0.01$, are visually in close agreement with the steady-state profiles. Note that $x = L$ is chosen as it is the location that takes the longest time to reach steady state.

In Table \ref{tab:results}, to quantitatively assess the accuracy of the finite transition time estimate for species $i$ at node $j$ we compute:
\begin{gather}
\label{eq:eps_i}
\varepsilon_{i,j} \approx \frac{c_{i,j}(\tau_{i,j}) - c_{i,\infty,j}}{c_{i,0}(x_{j}) - c_{i,\infty,j}},
\end{gather}
with a value of $\varepsilon_{i,j}$ close to $\delta$ indicating good accuracy of $\tau_{i,j}$ as evident from Eqs (\ref{eq:taui_definition})--(\ref{eq:fi}). The value of $c_{i,j}(\tau_{i,j})$ is computed using a finite volume scheme, briefly mentioned in Section \ref{sec:finite_transition_times}. The value of $c_{i,\infty,j}$ which approximates $c_{i,\infty}(x_{j})$ is computed using a similar spatial discretisation to that described for the moment boundary value problem (see Section \ref{sec:moments}). For full details on these solution strategies the reader is referred to our MATLAB code available on GitHub: \href{https://github.com/elliotcarr/Carr2020b}{https://github.com/elliotcarr/Carr2020b}. 

\begin{table*}[p]
\renewcommand*{\arraystretch}{0.7}
Case A\\[0.1cm]
\begin{tabular*}{0.95\textwidth}{@{\extracolsep{\fill}}ccrlrlrl}
\hline
$\delta$ & $k$ & $\tau_{1,N}$ & $\varepsilon_{1,N}$ & $\tau_{2,N}$ & $\varepsilon_{2,N}$ & $\tau_{3,N}$ & $\varepsilon_{3,N}$\\
\hline
$10^{-2}$ & 1 & 8.03 & \num{5.9070e-04} & 13.54 & \num{5.7309e-04} & 25.96 & \num{5.1595e-04}\\
& 5 & 5.30 & \num{1.0001e-02} & 9.00 & \num{1.0154e-02} & 17.11 & \num{1.0099e-02}\\
& 10 & 5.30 & \num{1.0001e-02} & 9.02 & \num{1.0006e-02} & 17.14 & \num{1.0021e-02}\\
& 15 & 5.30 & \num{9.9999e-03} & 9.05 & \num{9.8628e-03} & 17.16 & \cellcolor{tableshade}\num{9.9522e-03}\\
\hline
$10^{-4}$ & 1 & 16.07 & \num{1.4255e-07} & 27.07 & \num{1.0351e-07} & 51.91 & \num{8.2186e-08}\\
& 5 & 9.78 & \num{9.6418e-05} & 16.51 & \num{8.6691e-05} & 31.26 & \num{8.6517e-05}\\
& 10 & 9.75 & \num{1.0000e-04} & 16.28 & \num{1.0028e-04} & 30.82 & \num{1.0010e-04}\\
& 15 & 9.75 & \num{1.0000e-04} & 16.28 & \num{1.0007e-04} & 30.83 & \cellcolor{tableshade}\num{1.0002e-04}\\
\hline
$10^{-6}$ & 1 & 24.10 & \num{3.5020e-11} & 40.61 & \num{1.8921e-11} & 77.87 & \num{1.2595e-11}\\
& 5 & 14.26 & \num{9.2954e-07} & 24.01 & \num{7.2893e-07} & 45.40 & \num{7.3724e-07}\\
& 10 & 14.19 & \num{9.9987e-07} & 23.53 & \num{9.8997e-07} & 44.51 & \num{9.9456e-07}\\
& 15 & 14.19 & \num{1.0000e-06} & 23.51 & \num{1.0004e-06} & 44.50 & \cellcolor{tableshade}\num{1.0001e-06}\\
\hline
\end{tabular*}

\vspace*{0.1cm}
Case B\\[0.1cm]
\begin{tabular*}{0.95\textwidth}{@{\extracolsep{\fill}}ccrlrlrl}
\hline
$\delta$ & $k$ & $\tau_{1,N}$ & $\varepsilon_{1,N}$ & $\tau_{2,N}$ & $\varepsilon_{2,N}$ & $\tau_{3,N}$ & $\varepsilon_{3,N}$\\
\hline
$10^{-2}$ & 1 & 8.03 & \num{5.9070e-04} & 13.54 & \num{5.7309e-04} & 20.23 & \num{1.3410e-04}\\
& 5 & 5.30 & \num{1.0001e-02} & 9.00 & \num{1.0154e-02} & 12.22 & \num{1.0337e-02}\\
& 10 & 5.30 & \num{1.0001e-02} & 9.02 & \num{1.0006e-02} & 12.26 & \num{1.0150e-02}\\
& 15 & 5.30 & \num{9.9999e-03} & 9.05 & \num{9.8628e-03} & 12.57 & \cellcolor{tableshade}\num{8.6261e-03}\\
\hline
$10^{-4}$ & 1 & 16.07 & \num{1.4255e-07} & 27.07 & \num{1.0351e-07} & 40.45 & \num{1.2989e-09}\\
& 5 & 9.78 & \num{9.6418e-05} & 16.51 & \num{8.6691e-05} & 21.75 & \num{5.7365e-05}\\
& 10 & 9.75 & \num{1.0000e-04} & 16.28 & \num{1.0028e-04} & 20.71 & \num{1.0256e-04}\\
& 15 & 9.75 & \num{1.0000e-04} & 16.28 & \num{1.0007e-04} & 20.72 & \cellcolor{tableshade}\num{1.0204e-04}\\
\hline
$10^{-6}$ & 1 & 24.10 & \num{3.4964e-11} & 40.61 & \num{1.8929e-11} & 60.68 & \num{3.0568e-13}\\
& 5 & 14.26 & \num{9.2954e-07} & 24.01 & \num{7.2893e-07} & 31.27 & \num{2.5775e-07}\\
& 10 & 14.19 & \num{9.9987e-07} & 23.53 & \num{9.8997e-07} & 29.16 & \num{8.6367e-07}\\
& 15 & 14.19 & \num{1.0000e-06} & 23.51 & \num{1.0004e-06} & 28.87 & \cellcolor{tableshade}\num{1.0204e-06}\\
\hline
\end{tabular*}
\caption{Finite transition times for Cases A and B for different combinations of tolerance $\delta$ and moment index $k$. All finite transition times are calculated at the right boundary ($x = L$) via the formula (\ref{eq:tauixj}) with $j = N$. A value of $\varepsilon_{i,N}$ (\ref{eq:eps_i}) near $\delta$ indicates high accuracy of $\tau_{i,N}$. Shaded cells highlight the inferior accuracy of the finite transition time estimates for Case B.}
\label{tab:results}
\end{table*}

\begin{figure*}[t]
\includegraphics[width=0.32\textwidth]{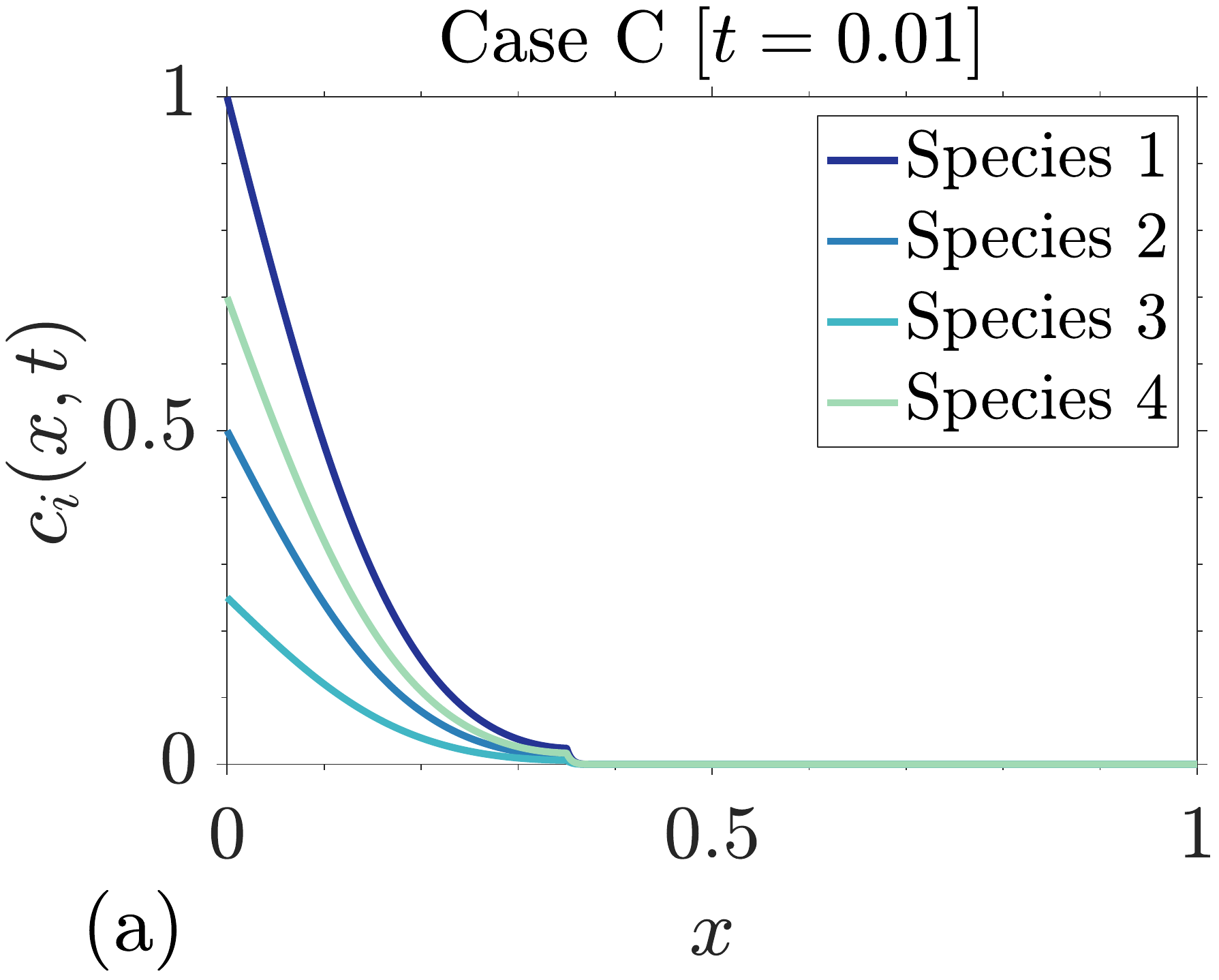}\hspace{0.01\textwidth}
\includegraphics[width=0.32\textwidth]{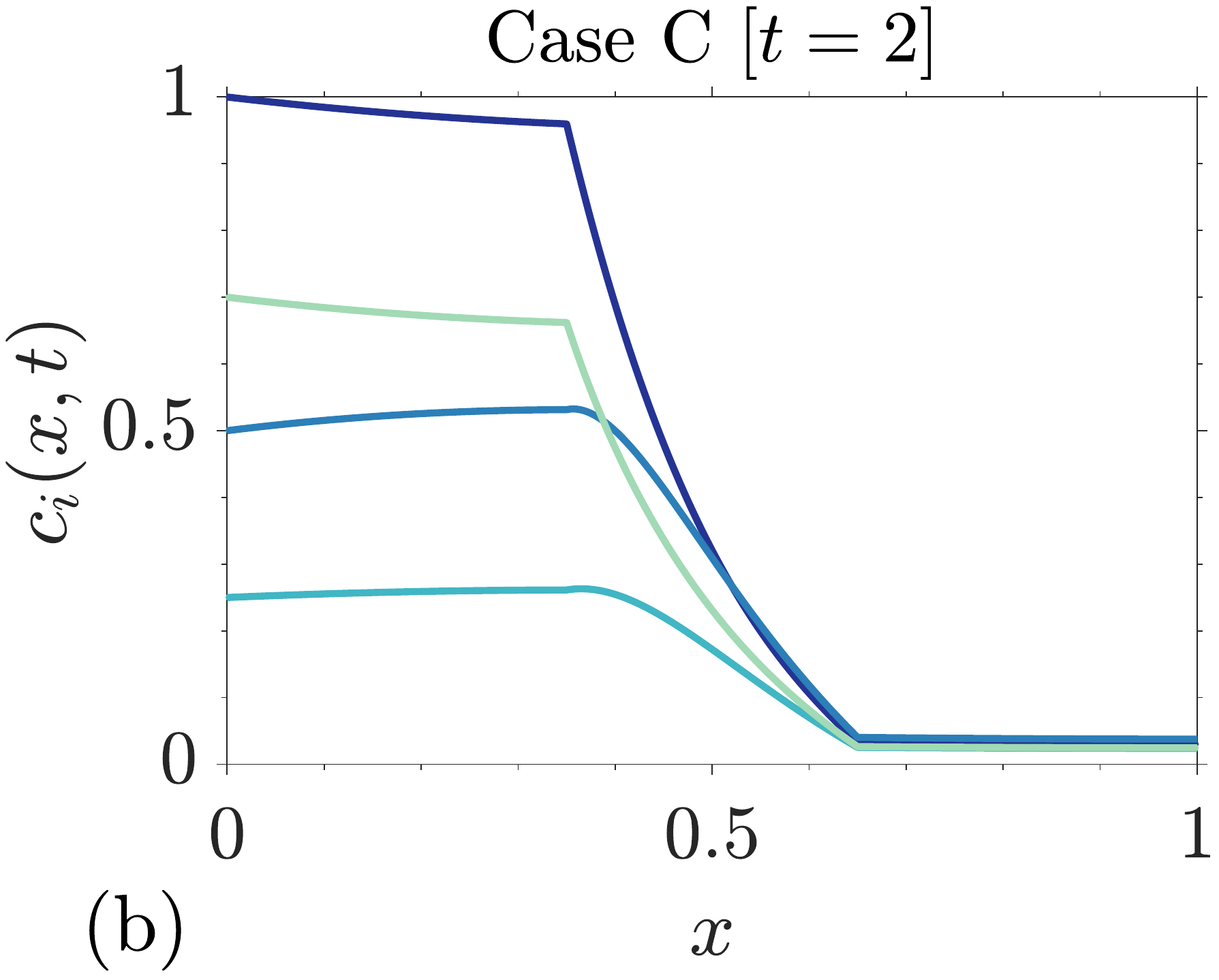}\hspace{0.01\textwidth}\includegraphics[width=0.32\textwidth]{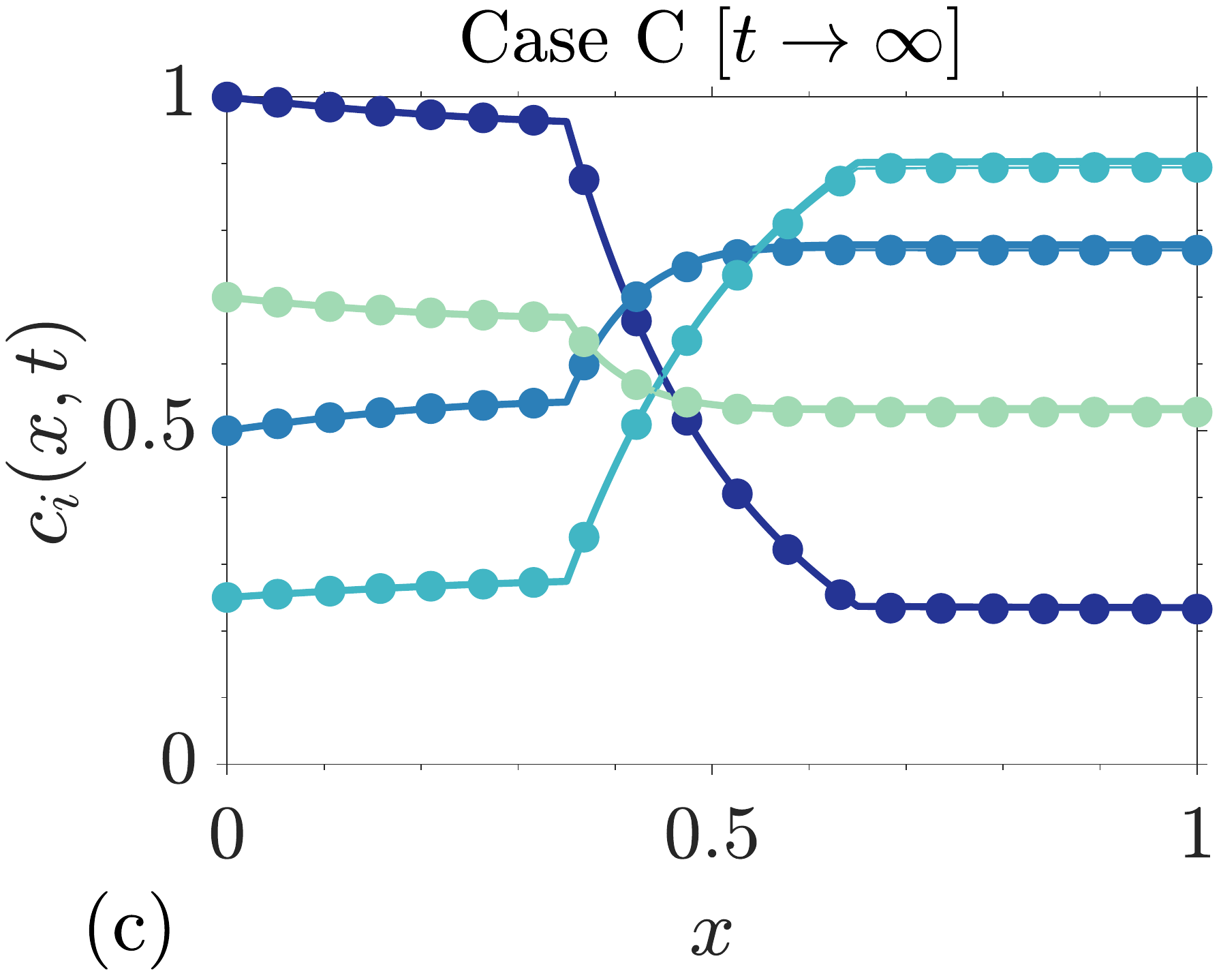}\\
\caption{\revision{Spatial distribution of the species concentration ($c_{i}(x,t)$, $i = 1,2,3,4$) over time for Case C. Circle markers in the $t\rightarrow\infty$ plot provide the concentration at the following finite transition times: $t = 60.08$ (species 1), $t = 65.08$ (species 2), $t = 66.70$ (species 3) and $t = 65.41$ (species 4). These finite transition times are calculated at the right boundary ($x = L$) via the formula (\ref{eq:tauixj}) with $j = N$, $\delta = 10^{-2}$ and $k = 15$. The accuracy of the estimates are verified in Table \ref{tab:results_C}. All solutions are obtained by solving the coupled reaction-diffusion model (\ref{eq:pde1})--(\ref{eq:bcR}) using the finite volume scheme discussed in Section \ref{sec:results} with $N = 501$ nodes.}}
\label{fig:results_C}
\end{figure*}

\begin{table*}[t]
\revision{
\renewcommand*{\arraystretch}{0.9}
Case C\\[0.1cm]
\begin{tabular*}{0.95\textwidth}{@{\extracolsep{\fill}}ccrlrlrlrl}
\hline
$\delta$ & $k$ & $\tau_{1,N}$ & $\varepsilon_{1,N}$ & $\tau_{2,N}$ & $\varepsilon_{2,N}$ & $\tau_{3,N}$ & $\varepsilon_{3,N}$ & $\tau_{4,N}$ & $\varepsilon_{4,N}$\\
\hline
$10^{-2}$ & 1 & 53.18 & \num{1.6439e-02} & 69.65 & \num{7.1923e-03} & 76.03 & \num{5.1032e-03} & 70.87 & \num{6.7489e-03}\\
& 5 & 60.08 & \num{9.9967e-03} & 65.08 & \num{9.9997e-03} & 66.70 & \num{1.0000e-02} & 65.41 & \num{9.9999e-03}\\
& 10 & 60.08 & \num{1.0000e-02} & 65.08 & \num{1.0000e-02} & 66.70 & \num{1.0000e-02} & 65.41 & \num{1.0000e-02}\\
& 15 & 60.08 & \num{1.0000e-02} & 65.08 & \num{1.0000e-02} & 66.70 & \num{1.0000e-02} & 65.41 & \num{1.0000e-02}\\
\hline
$10^{-4}$ & 1 & 106.37 & \num{3.5529e-04} & 139.30 & \num{4.7432e-05} & 152.06 & \num{2.1240e-05} & 141.73 & \num{4.0764e-05}\\
& 5 & 123.93 & \num{1.0015e-04} & 128.95 & \num{1.0002e-04} & 130.58 & \num{9.9953e-05} & 129.29 & \num{1.0000e-04}\\
& 10 & 123.95 & \num{1.0000e-04} & 128.95 & \num{1.0000e-04} & 130.58 & \num{1.0000e-04} & 129.29 & \num{1.0000e-04}\\
& 15 & 123.95 & \num{1.0000e-04} & 128.95 & \num{1.0000e-04} & 130.58 & \num{1.0000e-04} & 129.29 & \num{1.0000e-04}\\
\hline
$10^{-6}$ & 1 & 159.55 & \num{7.6785e-06} & 208.94 & \num{3.1280e-07} & 228.10 & \num{8.8496e-08} & 212.60 & \num{2.4624e-07}\\
& 5 & 187.78 & \num{1.0033e-06} & 192.82 & \num{1.0004e-06} & 194.46 & \num{9.9911e-07} & 193.16 & \num{1.0001e-06}\\
& 10 & 187.83 & \num{1.0000e-06} & 192.83 & \num{1.0000e-06} & 194.45 & \num{1.0001e-06} & 193.16 & \num{1.0000e-06}\\
& 15 & 187.83 & \num{1.0000e-06} & 192.83 & \num{1.0000e-06} & 194.45 & \num{1.0001e-06} & 193.16 & \num{1.0000e-06}\\
\hline
\end{tabular*}}
\caption{\revision{Finite transition times for Case C for different combinations of tolerance $\delta$ and moment index $k$. All finite transition times are calculated at the right boundary ($x = L$) via the formula (\ref{eq:tauixj}) with $j = N$. A value of $\varepsilon_{i,N}$ (\ref{eq:eps_i}) near $\delta$ indicates high accuracy of $\tau_{i,N}$.}}
\label{tab:results_C}
\end{table*}

Several observations are evident from Table \ref{tab:results}. Firstly, the accuracy improves as $k$ increases and $\delta$ decreases. This behaviour is consistent with that reported elsewhere \cite{carr_2017b,carr_2018a} and is due to the approximations (\ref{eq:lambda1})--(\ref{eq:alpha1}) improving for large $k$ and the approximation (\ref{eq:fi_larget}) improving for large $t$ (smaller values of $\delta$ produce larger values of $\tau_{i}$). Secondly, the accuracy of the finite transition time is inferior for Case B than Case A, for species 3 specificially (recall the only difference between Case A and B is the value of the reaction rate $\mu_{3}$). The reason for this is found by studying the matrix $\mathbf{A}$ in the differential system (\ref{eq:ode_system}), which has the following block structure:
\begin{gather*}
\mathbf{A} = \left[\begin{matrix} \widetilde{\mathbf{A}}+\mu_{1}\mathbf{I} &\\ 
-\mu_{1}\mathbf{I} & \widetilde{\mathbf{A}}+\mu_{2}\mathbf{I} & \\ 
& \ddots & \ddots \\
& & -\mu_{n-1}\mathbf{I} & \widetilde{\mathbf{A}}+\mu_{n}\mathbf{I} \end{matrix}\right],
\end{gather*}
where $\widetilde{\mathbf{A}}\in\mathbb{R}^{(N-1)\times(N-1)}$ is the matrix representing the discretised form of the operator $-\mathcal{L}$ (defined in Section \ref{sec:moments}) and $\mathbf{I}$ is the $(N-1)$ by $(N-1)$ identity matrix. Since $\mathbf{A}$ is block lower triangular, its eigenvalues ($\lambda_{j}$ for $j = 1,\hdots,\mathcal{N}$) are the union of the eigenvalues of the diagonal blocks: $\lambda_{j} = \xi_{k} + \mu_{i}$ for all $k = 1,\hdots,N-1$ and $i = 1,\hdots,n$, where $\xi_{k}$ ($k = 1,\hdots,N-1$) are the positive eigenvalues of $\widetilde{\mathbf{A}}$ ordered as $\xi_{1} < \xi_{2} < \cdots < \xi_{N-1}$. For Case B, $\mu_{3}$ is much closer to $\mu_{2}$ so the two smallest eigenvalues ($\lambda_{1} = \xi_{1}+\mu_{3}$ and $\lambda_{2} = \xi_{1}+\mu_{2}$, see Section \ref{sec:finite_transition_times}) are closer together and hence the assumption that $c_{i}(x,t)$ can be represented by a single exponential (\ref{eq:ci_larget}) for large $t$ is less valid. In summary, our approach works better \revision{for sequential reactions} if the two smallest values of $\mu_{i}$ ($i = 1,\hdots,n$) are well separated.

\begin{figure}
\includegraphics[width=0.45\textwidth]{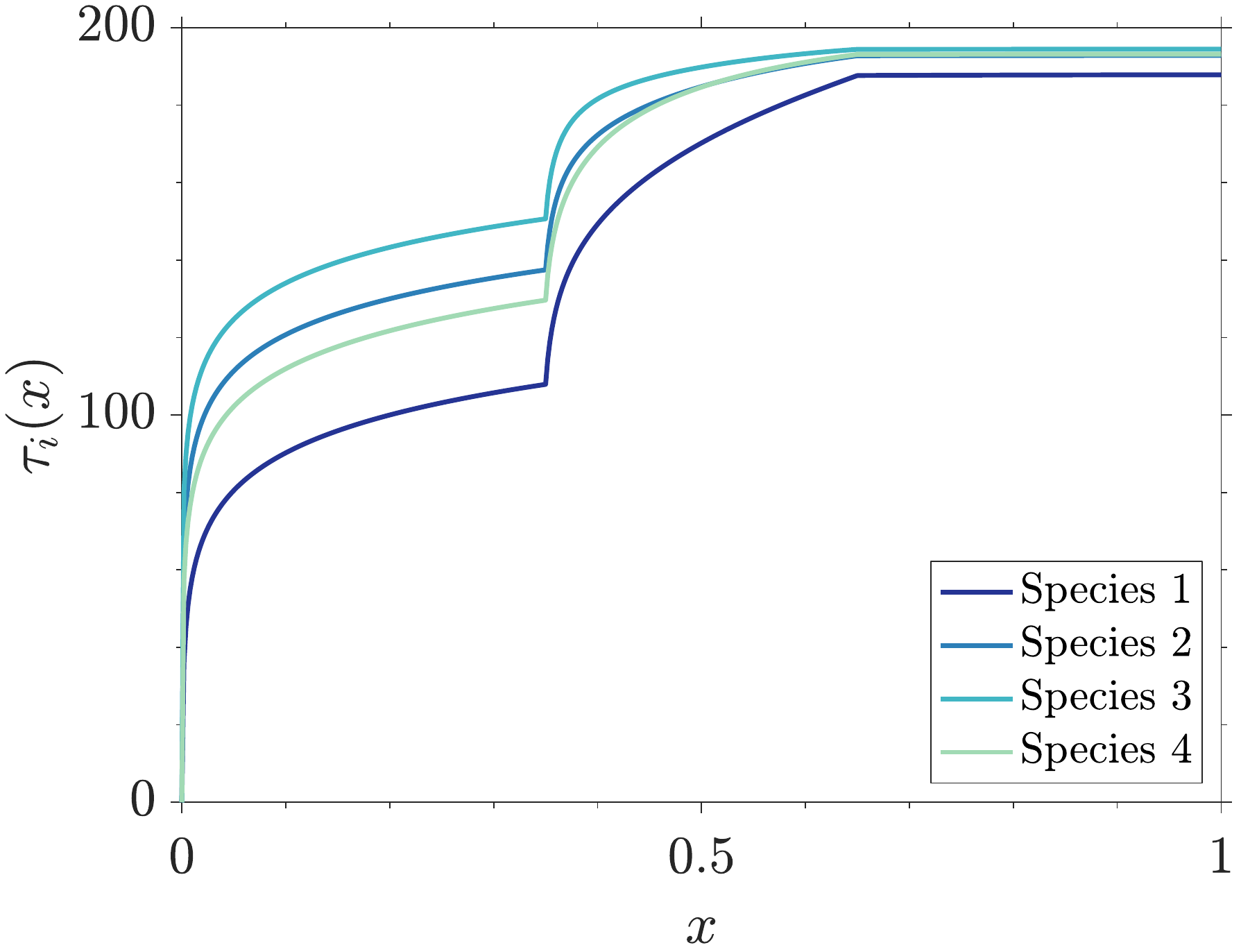}
\caption{\revision{Spatial distribution of the finite transition times ($\tau_{i}(x)$, $i = 1,2,3,4$) for Case C, calculated using the formula (\ref{eq:tauixj}) with $j = 1,\hdots,N$, $\delta = 10^{-6}$, $k = 15$ and $N = 501$ nodes. Note the values at $x = 1$ correspond to those given in Table \ref{tab:results_C} for $\delta = 10^{-6}$ and $k = 15$.}}
\label{fig:results_C_spatial}
\end{figure}

\revision{To further demonstrate the capability of our finite transition time formula, we present a third test case, Case C, involving four species and a more complex reaction network:
\begin{gather*}
\left[\begin{matrix*}[r] \mu_{1,1} & \mu_{1,2} & \mu_{1,3} & \mu_{1,4}\\ \mu_{2,1} & \mu_{2,2} & \mu_{2,3} & \mu_{2,4}\\ \mu_{3,1} & \mu_{3,2} & \mu_{3,3} & \mu_{3,4}\\ \mu_{4,1} & \mu_{4,2} & \mu_{4,3} & \mu_{4,4}\end{matrix*}\right] := \left[\begin{matrix*}[r] -0.4 & 0 & 0 & 0.12\\ 0.2 & -0.8 & 0 & 1.08\\ 0.1 & 0.24 & -0.2 & 0\\ 0.1 & 0.56 & 0.2 & -1.2\end{matrix*}\right],
\end{gather*}
where species $1$ produces species $2$, $3$ and $4$, species $2$ produces species $3$ and $4$, species $3$ produces species $4$ and species 4 produces species $1$ and $2$. The heterogeneous medium of length $L = 1$ is assumed to exhibit a layered structure  with diffusivity
\begin{gather*}
D(x) = \begin{cases} 1.0, & \text{if $0 \leq x \leq 0.35$,}\\ 0.01, & \text{if $0.35 < x < 0.65$,}\\ 1.0, & \text{if $0.65 \leq x \leq 1$.} \end{cases}
\end{gather*}
Non-zero concentrations are imposed for all four species at $x = 0$: $c_{b,1} = 1$, $c_{b,2} = 0.5$, $c_{b,3} = 0.25$ and $c_{b,4} = 0.7$. 

In Figure \ref{fig:results_C}, we plot the concentration profiles for each species at three times, depicting the transition from $t = 0$ to $t\rightarrow\infty$. In Figure \ref{fig:results_C}(c), the concentration profiles at the finite transition times, calculated at the right boundary ($x = L$) via the formula (\ref{eq:tauixj}) with $j = N$ and $\delta = 0.01$, are compared to the steady-state concentration profile. This comparison confirms that the calculated finite transition times provide an excellent finite measure of the time required for each species to reach steady state. Table \ref{tab:results_C} repeats the results for Cases A and B given in Table \ref{tab:results} and confirms for all four species the finite transition time formula is very accurate for $k > 5$.

For all the above results, the finite transition time formula is computed at the right boundary, $x = L$ (equivalently at node $j = N$). It is important to note, however, that the finite transition time is spatially dependent as evident from Eq (\ref{eq:tauixj}). In Figure \ref{fig:results_C_spatial}, for each species, we plot the spatial profile of the finite transition time for Case C, calculated at the right boundary ($x = L$) via the formula (\ref{eq:tauixj}) with $j = N$ and $\delta = 10^{-6}$. This figure demonstrates that the time required to effectively reach steady state for Case C monotonically increases. This makes intuitive sense since the larger the value of $x$, the further away from the left boundary ($x = 0$), which reaches steady-state instantaneously due to the identical Dirichlet boundary conditions (\ref{eq:bcL}) and (\ref{eq:bcL_ss}) shared by the transient model and it's steady-state analogue.}

\section{Conclusions}
\label{sec:conclusions}
This paper has presented new results for calculating how long a coupled heterogeneous reaction-diffusion process takes to effectively complete. Our approach extends recent analysis for single-species diffusion processes by demonstrating how the time taken can be accurately expressed in terms of higher-order moments of an appropriate function representing the transition of the process from initial to steady state. The attraction being that such moments can be calculated without having to solve the governing reaction-diffusion model. \revision{Three} presented test cases confirmed the efficacy of the approach and demonstrated that \revision{for sequential reactions} high accuracy is achieved if the two smallest reaction rates are well separated. 

\revision{Further validation of our approach could be achieved by comparing the finite transition time estimates to experimental breakthrough concentration curves \cite{gureghian_1985}. Experimental validation in this manner has been carried for a groundwater flow problem (diffusion with zero-order production/decay) in previous work by the first author \cite{carr_2018a}, where excellent agreement was reported between theory and experiment.} Finally, we remark that our analysis and our MATLAB code is limited to the coupled reaction-diffusion model described in Eqs (\ref{eq:pde1})--(\ref{eq:bcR}). More general initial and boundary conditions have been addressed previously for single-species diffusion processes \cite{carr_2017b} and these ideas carry over to the coupled multi-species reaction-diffusion processes considered in our work.


\begin{acknowledgments}
This research was partially supported by the Australian Mathematical
Sciences Institute (AMSI) who provided the second author with a 2019--2020 Vacation Research Scholarship. 
\end{acknowledgments}

\bibliographystyle{model1-num-names}
\bibliography{references}

\end{document}